\documentclass[11pt,a4paper]{article}
\usepackage{amsmath,amssymb,mathrsfs}
\usepackage[dvips]{graphicx}
\usepackage{epsfig}
\usepackage{comment}

\usepackage[toc]{appendix}

\def\un#1{\relax\ifmmode\@@underline#1\else
        $\@@underline{\hbox{#1}}$\relax\fi}

\let\du=\du                     

\def\a{\alpha}
\def\b{\beta}
\def\c{\chi}
\def\d{\delta}
\def\e{\epsilon}
\def\f{\phi}
\def\g{\gamma}
\def\h{\eta}

\def\l{\lambda}
\def\m{\mu}
\def\n{\nu}

\def\p{\pi}
\def\q{\theta}
\def\r{\rho}
\def\s{\sigma}
\def\t{\tau}

\def\x{\xi}

\def\F{\Phi}

\def\Q{\Theta}

\def\vf{\varphi}

\def\cc{{\cal C}}

\def\cg{{\cal F}}
\def\cg{{\cal G}}

\def\cl{{\cal L}}

\def\cp{{\cal P}}

\def\car{{\cal R}}
\def\cs{{\cal S}}

\def\bo{{\raise-.3ex\hbox{\large$\Box$}}}               
\def\pa{\partial}                                       
\def\TH{{\raise.2ex\hbox{$\displaystyle \bigodot$}\mskip-4.7mu \llap H \;}}
\def\face{{\raise.2ex\hbox{$\displaystyle \bigodot$}\mskip-2.2mu \llap {$\ddot
        \smile$}}}                                      

\def\VEV#1{\left\langle #1\right\rangle}        
\def\leftrightarrowfill{$\mathsurround=0pt \mathord\leftarrow \mkern-6mu
        \cleaders\hbox{$\mkern-2mu \mathord- \mkern-2mu$}\hfill
        \mkern-6mu \mathord\rightarrow$}
\def\dvec#1{\vbox{\ialign{##\crcr
        \leftrightarrowfill\crcr\noalign{\kern-1pt\nointerlineskip}
        $\hfil\displaystyle{#1}\hfil$\crcr}}}           
\def\dt#1{{\buildrel {\hbox{\LARGE .}} \over {#1}}}     

\def\sfrac#1#2{{\vphantom1\smash{\lower.5ex\hbox{\small$#1$}}\over
        \vphantom1\smash{\raise.4ex\hbox{\small$#2$}}}} 
\def\bfrac#1#2{{\vphantom1\smash{\lower.5ex\hbox{$#1$}}\over
        \vphantom1\smash{\raise.3ex\hbox{$#2$}}}}       
\def\afrac#1#2{{\vphantom1\smash{\lower.5ex\hbox{$#1$}}\over#2}}    

\def\[{\lfloor{\hskip 0.35pt}\!\!\!\lceil}
\def\]{\rfloor{\hskip 0.35pt}\!\!\!\rceil}
\def\Lag{{\cal L}}
\def\du#1#2{_{#1}{}^{#2}}

\def\ha{{\fracmm12}}

\def\un{\underline}
\def\fracmm#1#2{{{#1}\over{#2}}}

\def\low#1{{\raise -3pt\hbox{${\hskip 0.75pt}\!_{#1}$}}}

\def\Dot#1{\buildrel{_{_{\hskip 0.01in}\bullet}}\over{#1}}
\def\dt#1{\Dot{#1}}

\newskip\humongous \humongous=0pt plus 1000pt minus 1000pt

\newif\ifdtup

\newcommand{\be}{\begin{equation}}
\newcommand{\ee}{\end{equation}}
\newcommand{\nbe}{\begin{equation*}}
\newcommand{\nee}{\end{equation*}}

\newcommand{\fr}{\frac}
\newcommand{\lb}{\label}

\def\lessim{\lower0.6ex\hbox{$\,$\vbox{\offinterlineskip\hbox{$<$}\vskip1pt\hbox{$\sim$}}$\,$}}
\def\grtsim{\lower0.6ex\hbox{$\,$\vbox{\offinterlineskip\hbox{$>$}\vskip1pt\hbox{$\sim$}}$\,$}}

\def\ex{{\mathrm e}}

\numberwithin{equation}{section}

\begin{document}

\begin{titlepage}

\begin{center}

December 2015 \hfill IPMU15-0177\\
revised version

\noindent
\vskip2.0cm
{\huge \bf 

Starobinsky-like two-field inflation

}

\vglue.3in

{\large
Sho Kaneda~${}^{a}$ and Sergei V. Ketov~${}^{a,b,c}$ 
}

\vglue.1in

{\em
${}^a$~Department of Physics, Tokyo Metropolitan University \\
Minami-ohsawa 1-1, Hachioji-shi, Tokyo 192-0397, Japan \\
${}^b$~Kavli Institute for the Physics and Mathematics of the Universe (IPMU)
\\The University of Tokyo, Chiba 277-8568, Japan \\
${}^c$~Institute of Physics and Technology, Tomsk Polytechnic University\\
30 Lenin Ave., Tomsk 634050, Russian Federation \\
}

\vglue.1in
kaneda-sho@tmu.ac.jp, ketov@tmu.ac.jp

\end{center}

\vglue.3in

\begin{center}
{\Large\bf Abstract}
\end{center}
\vglue.1in

We consider an extension of the Starobinsky model, whose parameters are functions of an extra
scalar field. Our motivation is to test the robustness (or sensitivity) of the Starobinsky inflation against mixing scalaron with another (matter) scalar field. We find that the extended Starobinsky model is (classically) equivalent to the two-field inflation, with the scalar potential having a flat direction. For the sake of fully explicit calculations, we perform a numerical scan of the parameter space. Our findings support the viability of the Starobinsky-like two-field inflation for the certain range of its parameters, which is characterized by the scalar index $n_s=0.96\pm 0.01$, the tensor-to-scalar ratio $r<0.06$, and small running of the scalar index at $|\a_s|<0.05$.

\end{titlepage}


\section{Introduction}

Cosmological inflation in the early Universe is practically well established both theoretically and experimentally. It gives
the universal solution to many problems of the Standard Cosmology, because it predicts homogeneity of our Universe 
at large scales, its spatial flatness, its large size and entropy, as well as the almost scale-invariant spectrum of cosmological perturbations, in remarkable agreement with the COBE, WMAP, PLANCK and BICEP2 measurements of the {\it 
Cosmic Microwave Background} (CMB) radiation spectrum. Inflation is also thought of as the amplifier of microscopic quantum field fluctuations in vacuum, and it is the only known mechanism for seeds of the macroscopic structure formation. 

The standard mechanism of inflation in field theory uses a scalar field (called  inflaton), whose potential energy drives inflation. The inflaton scalar potential  should be flat enough to meet the slow-roll conditions during the inflationary stage. Physical
nature and fundamental origin of inflaton and its interactions to the {\it Standard Model}  (SM) elementary particles are unknown.
 
Starobinsky inflation \cite{star1,star2,mchi,star3,myrev} offers the gravitational origin of inflaton by identifying it with the 
spin-0 part of space-time metric. In the Higgs inflation \cite{hi1,hi2,ks} inflaton is identified with the Higgs field of the SM. Both those single-field inflationary models offer the very economic and viable descriptions of chaotic inflation together with the clear origin of the inflaton field either from gravitational theory or from particle theory, respectively. As regards slow-roll inflation, the predictions of the Starobinsky and Higgs inflationary models are essentially {\it the same} (see below in this Section).

The simplest Starobinsky model of inflation is based on the modified gravity action \cite{star1}
\be \lb{stara}
S[g] = \int \mathrm{d}^4x\sqrt{-g} \left[ -\ha R +\frac{1}{12M^2}R^2\right]
\ee
in terms of 4D spacetime metric $g_{\m\n}(x)$ with the scalar curvature $R$, where we have used the natural units
with the reduced Planck mass $M_{\rm Pl}=1$ and the space-time signature $(+,-,-,-)$.
 Slow-roll inflation takes place in the high-curvature regime 
($M\ll H\ll 1$ and $|\dt{H}|\ll H^2$), where the Hubble function $H(t)$ has been introduced. Then the
Starobinsky inflationary solution (attractor!) takes the simple form 
\be \lb{ssol}
H\approx \frac{M^2}{6}(t_{\rm exit}-t)~,\qquad 0<t\leq t_{\rm exit}~~.
\ee
The inflationary model (\ref{stara}) has a single mass parameter $M$ whose value is fixed by the observational Cosmic 
Microwave Background (CMB) data as $M=(3.0 \times10^{-6})(\fracmm{50}{N_e})$ where $N_e$ is the e-foldings number. The predictions of the Starobinsky model (\ref{stara}) for the spectral indices  $n_s\approx 1-2/N_e\approx 0.964$, $r\approx 12/N^2_e\approx0.004$ and low non-Gaussianity are in agreement with the WMAP and PLANCK 2013 data ($r<0.13$ and $r<0.11$, respectively, at 95\% CL) \cite{planck2}, though are in disagreement with the BICEP2 measurements ($r=0.2+0.07,-0.05$) \cite{bicep2}. The enhancement of the tensor-to-scalar-ratio $r$ of the Starobinsky model to the higher values can be achieved via modification of the simplest {\it Ansatz} (\ref{stara}) by (matter) quantum corrections (beyond one loop) \cite{kw11,des}.  However, the Planck 2015 data \cite{planck3} excludes a significant enhancement of $r$ beyond
$r=0.08$. Therefore, the Starobinsky model (\ref{stara}) still perfectly fits the current observational data.

It raises the natural question on the theory side about robustness of the simplest Starobinsky model (\ref{stara}) against mixing scalaron with other (matter) scalars. Though the current observational data favors a single-field inflation, it is very
unlikely that any single-field inflationary model is capable to provide the ultimate description of inflation.  As regards a more fundamental description of inflation in supergravity and string theories, multi-field inflation is a must, see e.g., Refs.~\cite{kt1,kt2,ktalk}.  The direct observational evidence for multi-field inflation would be a detection of primordial isocurvature
perturbations beyond the adiabatic spectrum (see Subsec.~3.3 for details). 

In this paper, we study the two-field extensions of the Starobinsky model by non-minimal couplings, motivated by generic supergravity extensions of Eq.~(\ref{stara}) in Ref.~\cite{kte}.

The action (\ref{stara}) can be dualized by the Legendre-Weyl transform \cite{lwtr,kkw}  to the standard (quintessence)  action of the Einstein gravity coupled to a single physical scalar (canonically normalized inflaton) $\f$ having the scalar potential
\be \lb{starp}
V(\f) = \fracmm{3}{4} M^2\left( 1- e^{-\sqrt{\frac{2}{3}}\f }\right)^2~.
\ee

During slow-roll inflation the $R^2$ term dominates in the action (\ref{stara}), whereas the coupling constant in front of it  is dimensionless. It implies the (approximate) rigid scaling invariance of the Starobinsky inflation in the high curvature $R$ (or in the large field $\f \to +\infty$) limit \cite{ks}. The scaling invariance is not exact for finite values of $R$, and its violation is exactly measured by the slow-roll parameters, in full correspondence to the observed (nearly conformal) spectrum of the CMB perturbations. The (approximate) flatness of the inflaton scalar potential implies the (approximate) shift symmetry of the inflaton field. It also implies the alternative physical interpretation of the inflaton field as the pseudo-Nambu-Goldstone boson associated
with spontaneous breaking of the scale invariance \cite{nat1,nat2,nat3}.

Similar observations apply to the Higgs inflation in the presence of a {\it non-minimal} coupling of the Higgs field to the space-time scalar curvature \cite{hi1}. It also has the approximate (rigid) scale invariance and, actually, {\it the same} scalar potential (\ref{starp}) during slow roll inflation \cite{ks}. The Higgs inflation is based on the Lagrangian  (in the Jordan frame) \cite{hi1}
\be \lb{bl}
\Lag_{\rm J} = \sqrt{-g_{\rm J}}\left\{ -\ha (1+\x\f_{\rm J}^2)R_{\rm J} + 
\ha g^{\m\n}_{\rm J}\pa_{\m}\f_{\rm J}\pa_{\n}\f_{\rm J} -V_{\rm H}(\f_{\rm J}) \right\}~~,
\ee
having the real scalar field $\f_{\rm J}(x)$ non-minimally coupled to gravity with the coupling constant $\x$, and 
the  Higgs scalar potential  
\be \lb{hpot}
V_{\rm H}(\f_{\rm J}) = \fracmm{\l}{4}(\f_{\rm J}^2-v^2)^2~~.
\ee

The action (\ref{bl}) can be rewritten to the Einstein frame by the  Weyl transformation
\be \lb{me}
g^{\m\n} = \fracmm{g_{\rm J}^{\m\n}}{(1+\x\f_{\rm J}^2)}~~.
\ee
It gives rise to the standard Einstein-Hilbert term $(-\ha R)$ for gravity in
the Lagrangian. However, it also leads to a non-minimal (or non-canonical) 
kinetic term of the scalar field $\f_{\rm J}$. To get the canonical kinetic term,
a scalar field redefinition is needed, $\f_{\rm J}\to \vf(\f_{\rm J})$, 
subject  to the condition
\be \lb{fre}
 \fracmm{d\vf}{d\f_{\rm J}} = \fracmm{ 
\sqrt{1+\x(1+6\x)\f_{\rm J}^2}}{1+\x\f_{\rm J}^2}~~.
\ee
As a result, the non-minimal theory (\ref{bl}) is classically equivalent to
the standard (canonical) theory of the scalar field $\vf(x)$ minimally coupled 
to gravity,
\be \lb{mina}
\Lag_{\rm E} = \sqrt{-g}\left\{ -\ha R + \ha g^{\m\n}\pa_{\m}\vf\pa_{\n}\vf
 -V(\vf) \right\}~~,
\ee
and having the scalar potential \cite{hi1}
\be \lb{hpote}
V(\vf) = \fracmm{V_{\rm H}(\f_{\rm J}(\vf))}{[1+\x\f_{\rm J}^2(\vf)]^2}~~.
\ee
Given a large positive $\x\gg 1$, one easily finds in the large field limit 
$\vf\gg\sqrt{\fracmm{2}{3}}\x^{-1}$ that
\be \lb{fres}
 \vf \approx \sqrt{\fracmm{3}{2}} \log (1+\x\f_{\rm J}^2)
\ee
and
\be \lb{lp}
V(\vf) \approx \fracmm{\l}{4\x^2}\left(
1-\exp\left[ -\sqrt{\fracmm{2}{3}} \vf\right] \right)^2
\ee
indeed. Comparing Eqs.~(\ref{starp}) and (\ref{lp}) gives rise to the identification \cite{ks}
\be \lb{freq}
M = \sqrt{\fracmm{\l}{3}}\,\x^{-1}~.
\ee  

The LHC and TEVATRON measurements of the masses of Higgs and t-quark, however, imply (via the renormalization group and
the Standard Model particle content) that the effective Higgs potential coupling constant $\l$ becomes negative  at around
$10^{11}~GeV$ \cite{smo}, which is lower than the expected scale of inflation. It means that the SM has to be extended by new particles and new physics.

It is still possible that inflaton is neither Starobinsky scalaron nor Higgs field, but a mixture of them. This possibility leads to a {\it two-field} inflation also. Another motivation to study the Starobinsky-like two-field inflation comes from 4D, N=1 supergravity with chiral matter superfields, where inflaton is automatically extended to a complex field as the leading bosonic field component of an N=1 scalar supermultiplet. For example, as was demonstrated in Ref.~\cite{kte}, a generic N=1 supergravity extension of the simplest Starobinsky model (\ref{stara}) leads to the non-minimal couplings of the Higgs field to both $R$ {\it and} $R^2$ gravity terms. And it is commonplace in string cosmology that inflaton is mixed with other scalars (moduli), so that a stabilization of the latter is required for inflation.

Our paper is organized as follows. In Sec.~2 we define the new class of two-field inflationary models as a combination (and a generalization) of Eqs.~(\ref{stara}) and (\ref{bl}), and rewrite them to the more standard (dual) form. Those inflationary models interpolate between the Starobinsky and Higgs (single-field) inflationary models, and can accommodate a broader range of values for the tensor-to-scalar ratio. In Sec.~3 we set up the equations of motion, and
classify our model against the other two-field inflationary models studied in the literature. In Sec.~4 we focus on the particular case by dropping the Higgs part of the scalar potential. In Sec.~5 we summarize our numerical findings in the special two-field model of the  Starobinsky-like inflation. Sec.~6 is our Conclusion. The technical details about linear perturbations, their spectra and evolution are collected in Appendix A.

\section{Starobinsky- and Higgs-inspired two-field inflation models}

Our new inflationary model (in Jordan frame) of a real scalar field $\f$ non-minimally coupled to the Starobinsky $(R+R^2)$ gravity is given by 
\begin{equation} \lb{ourm1}
(-g)^{-1/2}\cl = -\frac{1}{2}f^2(\f)R + \frac{1}{12M^2(\f)}R^2 +\frac{1}{2}g^{\m\n}\pa_{\m}\f \pa_{\n}\f -V(\f)~~.
\end{equation}
Its non-minimal couplings are described by {\it two} generic functions $f(\f)$ and $M(\f)$ in place of the constant parameters $M_{\rm Pl}$ and $M$ of the original Starobinsky model (\ref{stara}).~\footnote{More general couplings,
including an arbitrary function of $\f$ {\it and} $R$, were considered in Ref.~\cite{vag}.}

 Both functions enter the Lagrangian
 (\ref{ourm1}) via their squares, in order to avoid ghosts. It is worth mentioning that  both non-minimal couplings are required by renormalization of the $(R+R^2)$ gravity coupled to matter. In other words, we just replaced the parameters of the Starobinsky  $(R+R^2)$ gravity by functions of a (Higgs) scalar field $\f$.

Should the scalar field $\f$ be stabilized by its scalar potential $V(\f)$ to some vacuum expectation value $\f_0$, 
our model reduces to the standard Starobinsky model (Sec.~1). Should the $M^2(\f)$ be sent to infinity, the Higgs
inflationary setup is recovered. 
 
 In the case of the truly Higgs field $\f$, its scalar potential takes the form
\be \lb{higgs}
V_{\rm H}(\f) = \fr{\l}{4}\left(\f^2 - v^2 \right)^2
\ee
in terms of the coupling constants $\l>0$ and $v=\VEV{\f}_0\equiv \f_0$. 

Thus, the model (\ref{ourm1}) describes all the quintessence models with a non-minimal coupling to $R$ (like the Higgs inflation) and 
the $R+R^2$ gravity model of Starobinsky (\ref{stara}) as the particular cases. A non-minimal coupling to the $R^2$
term is our new feature when $M(\f)$ is truly field-dependent.

In order to understand the physical significance of our model, and put it under the standard treatment  in theoretical cosmology
(in Einstein frame), let us replace the $R^2$ by the $2\c R - {\c}^2$ in Eq.~(\ref{ourm1}), where the new scalar $\c$ has been
introduced as follows:
\begin{equation} \lb{ourm2}
(-g)^{-1/2}\cl= -\frac{1}{2} \left( f_1^2 - \frac{1}{3M^2}\c \right) R -\frac{1}{12M^2} \c^2 + 
\frac{1}{2}g^{\m\n}\pa_{\m}\f \pa_{\n}\f -V_{\rm H}(\f)~.
\end{equation}
It is easy to check that it is classically equivalent to the original model (\ref{ourm1}) because the equation of
motion of the $\c$ field is algebraic, and its solution reads $\c=R$.

Introducing the notation 
\be \lb{not1}
A(\f, \c) = f_1^2(\f) - \frac{1}{3M^2(\f)}\c
\ee
allows us to rewrite Eq.~(\ref{ourm2}) to the Brans-Dicke-type form
\begin{equation} \lb{ourm3}
(-g)^{-1/2}\cl = -\frac{1}{2}AR - \frac{1}{12M^2}\c^2 + \frac{1}{2}g^{\m\n}\pa_{\m}\f\pa_{\n}\f - V_{\rm H}(\f)~.
\end{equation}

When assuming positivity of $A$ (to avoid ghosts), i.e.
\begin{equation} \lb{nogc}
f_1^2(\f) > \frac{1}{3M^2}\c~,
\end{equation}
the Weyl transformation of metric, $g_{\m\n} \rightarrow A(\f, \c)g_{\m\n}$, gives rise
to the standard (Einstein-Hilbert) term for gravity in the classically equivalent (dual) Lagrangian,
\begin{equation} \lb{dualla}
(-g)^{-1/2}\cl = - \frac{1}{2} R + \frac{3}{4A^2}g^{\m\n}\pa_{\m}A\pa_{\n}A  + \frac{1}{2A}g^{\m\n}\pa_{\m}\f\pa_{\n}\f - 
\frac{1}{12M^2}\c^2 -A^2V_{\rm H}(\f)~,
\end{equation}
where we have used the Einstein-frame metric and have ignored an additive total derivative. Hence, our model takes the form of the {\it Non-Linear Sigma-Model} (NLSM) \cite{book}
\be \lb{nlsm}
(-g)^{-1/2}\cl = - \frac{1}{2} R + \frac{1}{2}g^{\m\n}G_{ij}(\f,\c)\pa_{\m}\f_i \pa_{\n}\f_j  - V(\f,\c)~,
\ee
having the NLSM metric ($i,j=1,2$, the primes denote the derivatives with respect to $\f$)
\begin{equation} \lb{nm}
G_{ij} =\frac{1}{A^2} 
\left(
\begin{array}{cc}
  \frac{3}{2}\left(2f_1f_1^{\prime} + \frac{2}{3}M^{-3}M^{\prime}\c \right)^2 + A  &  -M^{-2} \left(f_1f_1^{\prime} + \frac{1}{3}M^{-3}M^{\prime}\c \right)   \\
  -M^{-2} \left(f_1f_1^{\prime} + \frac{1}{3}M^{-3}M^{\prime}\c \right)   &   \frac{3}{2} \left(\frac{1}{3}M^{-2} \right)^2\\ 
\end{array}
\right)
\end{equation}
in terms of the two scalars $\f_{1,2} =(\f, \c)$, minimally coupled to gravity in the physical (Einstein) frame, and having the
scalar potential
\be \lb{twospot}
V(\f,\c) = \frac{1}{12M^2}\c^2 +A^2V_{\rm H}(\f)~.
\ee

The NLSM kinetic terms have no ghosts under the condition (\ref{nogc}) because
\begin{equation} \lb{det}
\det G_{ij} = \frac{1}{6M^4A^3}~.
\end{equation}

The NLSM in Eqs.~(\ref{nlsm}) and (\ref{nm}) can be further simplified by considering $A$ as the new independent scalar field (instead of $\c$)  and doing the field redefinition
\be \lb{fred}
 A= \exp \left( \sqrt{ \frac{2}{3}} \ \r \right)~,
\ee
which leads to the {\it canonical} kinetic term of the scalar field $\r$. We find
\be \lb{mod}
\begin{split}
(-g)^{-1/2}\cl =  & -\fr{1}{2}R  +\fr{1}{2}g^{\m\n} \pa_{\m}\r \pa_{\n}\r  
+\fr{1}{2} \mathrm{exp}\left( -\sqrt{\fr{2}{3}}\r \right) g^{\m\n} \pa_{\m}\f \pa_{\n}\f \\
              & -\fr{3}{4}M^2(\f) \left[ f^2(\f) -\mathrm{exp} \left( \sqrt{\fr{2}{3}}\r \right) \right]^2 - 
              \exp \left( 2 \sqrt{ \frac{2}{3}} \ \r \right)V_{\rm H}(\f)~. \\
\end{split}
\ee

Hence, the full scalar potential $V(\f,\r)$ is given by
\be \lb{fullsp}
V(\f,\r)=  \fr{3}{4}M^2(\f) \left\{ f^2(\f) -\mathrm{exp} \left( \sqrt{\fr{2}{3}}\r \right) \right\}^2 + \exp \left( 2\sqrt{ \frac{2}{3}} \ \r \right)V_{\rm H}(\f)~.
\ee

The first term of the scalar potential has a {\it flat direction} along
\be \lb{flat}
\r =  \sqrt{\fr{3}{2}}\ln f^2(\f)~,
\ee
while the second term in the case (\ref{higgs}) has the absolute minimum at $\f=v$. Therefore, along the flat direction
(\ref{flat}), the full scalar potential  (\ref{fullsp}) has the absolute minimum at $\f=v$ in the Minkowski vacuum. Moreover, along the flat direction (\ref{flat}), our model reduces to a single-field model having the scalar potential 
$V(\f)$ in terms of the non-canonical scalar $\f$. The latter can be traded for a canonical scalar by a field redefinition, similarly to that of Eq.~(\ref{fre}). 

We would like to emphasize that the discovered existence of a flat direction is automatic in the class of models under consideration, and it does not require supersymmetry. In a generic solution, two scalar fields $\f$ and $\r$ are going to evolve towards the flat direction.

In order to make our model to be more specific and treatable for numerical calculations, in what follows we eliminate the functional freedom above by choosing 
\be \lb{para}
f^2(\f) = 1 + \a \f^2 \qquad {\rm and} \qquad M^2(\f)= M^2(1  +\b \f^2)
\ee
with some (non-negative) coupling constants $\a$, $\b$ and $M$. 

The choice (\ref{para}) is also motivated by renormalizability. Though each of the quantum field theories (\ref{ourm1}) and (\ref{mod}) is not renormalizable as a theory of quantum gravity, it still makes sense to demand the (limited) renormalizability of the quantized scalar sector in a classical (curved) gravitational background. Then the non-minimal couplings (\ref{para}) naturally arise with the renormalization counterterms \cite{bd,bos}. The Higgs potential (\ref{higgs}) also fits the limited renormalizability requirement.

The field theory  (\ref{mod}) has two real scalars $(\r,\f)$ minimally coupled to the Einstein gravity and having
the scalar potential  (\ref{fullsp}). The kinetic term of the $\r$ scalar is canonically normalized, whereas the canonical
term of the $\f$ scalar has the $\r$-dependent factor. In the next Sections we study two-field inflation in those models.
\vglue.2in

\section{Classification of our model against the literature}

Though our model  (\ref{mod})  has the non-canonical kinetic term of $\f$, it falls into the class of the two-field inflationary models studied, e.g., in Ref.~\cite{lal}  in the slow roll approximation and having the action

\be
S = \int d^4x \sqrt{-g} \left[- \fr{1}{2}R  +\fr{1}{2}(\pa_{\m}\r)(\pa^{\m}\r) +\fr{\ex^{2b(\r)}}{2}(\pa_{\m}\f)(\pa^{\m}\f) - V(\r, \f) \right]~.
\label{eq1}
\ee

As regards cosmological perturbations, their spectra and evolution in the theory (\ref{eq1}),  we employ the results of Refs.~\cite{lal,jap,ktsu,uk,uk2} in the linear approximation with respect to the slow-roll parameters. To make
our paper self-contained and complete, a derivation of the relevant equations is summarized in Appendix A.  Those results
are used in Sec.~5 for our numerical analysis.

\subsection{Equations of motion}

When assuming a spatially flat {\it Friedmann-Lemaitre-Robertson-Walker} (FLRW) universe with the metric
\be
ds^2 =  dt^2 - a(t)^2 d\mathbf{x}^2~,
\label{eq2}
\ee
the field equations in the theory (\ref{eq1}) take the form
\be
\ddot{\r} +3H\dot{\r} + V_{,\r} = b_{,\r} \ex^{2b} \dot{\f}^2~,
\label{eq3}
\ee
\be
\ddot{\f} + (3H + 2b_{,\r}\dot{\r}) \dot{\f} + \ex^{-2b} V_{,\f} = 0~,
\label{eq4}
\ee
\be
H^2 = \fr{1}{3} \left[ \fr{1}{2}\dot{\r}^2 + \fr{1}{2}\ex^{2b}\dot{\f}^2 +V \right]~,
\label{eq5}
\ee
\be
\dot{H} = - \fr{1}{2} \left[ \dot{\r}^2 + \ex^{2b}\dot{\f}^2 \right]~,
\label{eq6}
\ee
where the dots stand for the time derivatives, the subscripts after the comma denote the derivatives with respect to the
fields, and $H=\dt{a}/a$ is the Hubble function. 

In two-field inflation the slow-roll parameters form $2\times 2$ {\it matrices}, $(I,J=(\r,\f)=1,2)$,
\be
\e\low{IJ} \equiv \left(
\begin{array}{cc}
\e_{\r\r}& \e_{\r\f} \\
\e_{\r\f}& \e_{\f\f} \\
\end{array}
\right)
\ \ \ \ \ and \ \ \ \ \ 
\h\low{IJ} \equiv \left(
\begin{array}{cc}
\h_{\r\r}& \h_{\r\f} \\
\h_{\r\f}& \h_{\f\f} \\
\end{array}
\right)~~,
\label{eq9-1}
\ee
whose entries are given by
\be
\e_{\r\r} = \fr{\dot{\r}^2}{2H^2},\ \ \ \e_{\r\f} = \ex^{b}\fr{\dot{\r}\dot{\f}}{2H^2},\ \ \ \e_{\f\f} = \ex^{2b}\fr{\dot{\f}^2}{2H^2}~,
\label{eq7}
\ee
\be
\h_{IJ} = \fr{V_{,IJ}}{3H^2}~.
\label{eq8}
\ee
\be
\e \equiv \e_{\r\r} + \e_{\f\f} = -\fr{\dot{H}}{H^2}~.\lb{eq8-1}
\ee

In the case (\ref{eq1}) one finds
\be
\e\low{IJ} = \fr{1}{2H^2} \left( 
\begin{array}{cc}
\dot{\r}^2 & \ex^{b}\dot{\r}{\f} \\
\ex^{b}\dot{\r}{\f} & \ex^{2b}\dot{\f}^2 \\
\end{array}
\right)
\label{eq9-2}
\ee
and
\be
\h\low{IJ} = \fr{1}{3H^2} \left(
\begin{array}{cc}
V_{,\r\r}& V_{,\r\f}\\
V_{,\r\f}& V_{,\f\f}\\
\end{array}
\right)~.
\label{eq9-3}
\ee

\vglue.2in

\subsection{Correspondence of our model to the literature}

In our case (\ref{mod}), we have to specify above that
\be \lb{bfun}
b(\r) = -\fr{1}{2}\sqrt{\fr{2}{3}}\r~, \quad b_{,\r} = -\fr{1}{2}\sqrt{\fr{2}{3}}~,\quad b_{,\r\r} = 0~,
\ee
as well as
\be
V_{,\r} = \sqrt{\fr{3}{2}} M^2 \mathrm{exp}\left(\sqrt{\fr{2}{3}}\r\right) \left[ \mathrm{exp}\left(\sqrt{\fr{2}{3}}\r\right) - f^2 \right]~,
\ee
\be
V_{,\f} = \fr{3}{2}MM_{,\f}\left[ f^2 - \mathrm{exp}\left(\sqrt{\fr{2}{3}}\r\right) \right]^2 + \fr{3}{2}M^2f_{,\f} \left[2f^3 
-2f\mathrm{exp}\left(\sqrt{\fr{2}{3}}\r\right)\right] +(V_{\rm H})_{\f}~,
\ee
and 
\be
V_{,\r\r} = M^2 \mathrm{exp}\left(\sqrt{\fr{2}{3}}\r\right) \left[ 2\mathrm{exp}\left(\sqrt{\fr{2}{3}}\r\right)-f^2 \right]~,
\ee
\be
V_{,\r\f} = \sqrt{6}M \mathrm{exp}\left(\sqrt{\fr{2}{3}}\r\right) \left[ M_{,\f} \left\{ \mathrm{exp}\left(\sqrt{\fr{2}{3}}\r\right) - f^2 \right\} 
-Mff_{,\f} \right]~,
\ee
\be
\begin{split}
V_{,\f\f} = & \fr{3}{2} \left(M^2_{,\f} + MM_{,\f\f}\right) \left[ f^2 -\mathrm{exp}\left(\sqrt{\fr{2}{3}}\r \right) \right]^2 \\
                & +\fr{3}{2} \left( 4MM_{,\f}f_{,\f} + M^2 f_{,\f\f} \right) \left[ 2f^3 - 2f\mathrm{exp}\left(\sqrt{\fr{2}{3}}\r \right) \right] \\
                & +\fr{3}{2} M^2 f^2_{,\f} \left[ 6f^2 -2\mathrm{exp}\left(\sqrt{\fr{2}{3}}\r \right) \right] + (V_{\rm H})_{,\f\f}~,
\end{split}
\ee
where we have used the scalar potential (\ref{fullsp}).

The equations of motion for cosmological perturbations are given by Eqs.~(\ref{EOM for pb1}) and (\ref{EOM for pb2}),  whose  "coefficients" are listed in Eqs.~(\ref{eq37}), (\ref{eq38}), (\ref{eq39}) and  (\ref{eq40}) of Appendix A. Therefore (see Appendix A also), the perturbation (power) spectra are given by Eqs.~(\ref{ps si}), (\ref{ps s}) and (\ref{cor}) at the horizon crossing, and by Eqs.~(\ref{eq89}), (\ref{eq90}) and (\ref{eq91}) on super-Hubble scales, with
\be
\x = -\sqrt{\fr{\e}{3}}~.
\label{eq115}
\ee 

\section{Special case with $V_{\rm H}=0$ }

A simple two-field inflationary model of the same type defined in our Eq.~(\ref{ourm1}), though with a mass term instead of the Higgs scalar potential and {\it without} non-minimal interactions to $R$ or $R^2$, was considered in Ref.~\cite{uk2}. It was found by numerical calculations in Ref.~\cite{uk2} that the Starobinsky inflation is {\it robust} against that extension for a certain range of the ratio of two scalar masses. The model of Ref.~\cite{uk2} is in good agreement with the Planck data  \cite{planck3}. Multi-field dynamics of Higgs inflation in the presence of non-minimal couplings was analyzed
in Ref.~\cite{k2} where it was found that it is also in very good agreement with the Planck measurements.

In what follows, we restrict ourselves to the {\it different case} with $V_{\rm H}=0$ (or in the limit $\l\to 0$), when the
functions $f(\f)$ and $M(\f)$ are given by Eq.~(\ref{para}), for simplicity.  Then the  scalar potential reads
\be \lb{oursp}
V(\f,\r) = \fr{3}{4} M^2\left( 1+\b\f^2 \right) \left[ \left(1+\a\f^2 \right) - \exp\left(\sqrt{\fr{2}{3}}\r\right) \right]^2~.
\ee

The slow-roll matrices take the form
\be
\e_{IJ} = \fr{1}{2H^2} \left( 
\begin{array}{cc}
\dot{\r}^2 & \ex^{-\fr{1}{2}\sqrt{\fr{2}{3}}\r}\dot{\r}\dot{\f} \\
\ex^{-\fr{1}{2}\sqrt{\fr{2}{3}}\r}\dot{\r}\dot{\f} & \ex^{-\sqrt{\fr{2}{3}}\r}\dot{\f}^2 \\
\end{array}
\right)~,
\label{eq117}
\ee
whereas the $\h_{IJ}$ is the same as that in Eq.~(\ref{eq9-3})  with 
\be
M^{-2}V_{,\r\r} = \left(1+ \b\f^2\right) \exp\left(\sqrt{\fr{2}{3}}\r\right) \left[2\exp\left(\sqrt{\fr{2}{3}}\r\right) - (1+\a\f^2) \right],
\label{eq119}
\ee
\be
M^{-2}V_{,\r\f} = \sqrt{6}\b\f \exp\left(\sqrt{\fr{2}{3}}\r\right) \left[\exp\left(\sqrt{\fr{2}{3}}\r\right) - \left(1+\a\f^2\right)\right] -\sqrt{6}\a\f \left(1+\b\f^2\right)\exp\left(\sqrt{\fr{2}{3}}\r\right),
\label{eq120}
\ee
\be
\begin{split}
M^{-2}V_{,\f\f} =& \fr{3}{2}\b \left[\left(1+\a\f^2\right)-\exp\left(\sqrt{\fr{2}{3}}\r\right)\right]^2 \\
                 & +12\a\b\f^2\left(1+\a\f^2\right) +3\a\left(1+\a\f^2\right)\left(1+\b\f^2\right) -3\a^2\f^2\left(1+\b\f^2\right) \\
                 & -12\a\b\f^2\exp\left(\sqrt{\fr{2}{3}}\r\right) -3\a\left(1+\b\f^2\right)\exp\left(\sqrt{\fr{2}{3}}\r\right) \\
                 & +3\a^2\f^2\left(1+\a\f^2\right)^{-1}\left(1+\b\f^2\right)\exp\left(\sqrt{\fr{2}{3}}\r\right) \\
                 & +9\a^2\f^2\left(1+\b\f^2\right) -3\a^2\f^2\left(1+\a\f^2\right)^{-1}\left(1+\b\f^2\right)\exp\left(\sqrt{\fr{2}{3}}
                 \r\right)~.\\ 
\end{split}
\label{eq121}
\ee

In particular, the Hubble function squared of  Eq.~(\ref{eq5}) reads as 
\be
H^2 = \fr{1}{6}\dot{\r}^2 + \fr{1}{6}\ex^{-\sqrt{\fr{2}{3}}\r} \dot{\f}^2 + \fr{1}{4}\left(1+\b\f^2\right) \left[ (1+\a\f^2) -\exp \left(\sqrt{\fr{2}{3}}\r\right) \right]^2~.
\label{eq122}
\ee

\section{Numerical results}

The two-field scalar potential (\ref{oursp}) is semi-positively definite. It reduces to the Starobinsky scalar potential (\ref{starp}) at a fixed (or stabilized) $\f$, and to a power-law scalar potential $V/M^2$ (as a sum of the $\a^2\f^4$ and the $\a^2\b\f^6$ terms) at a fixed (or stabilized)  $\r$. 

\begin{figure}[ht] 
\begin{center}
\includegraphics[clip,width=8cm]{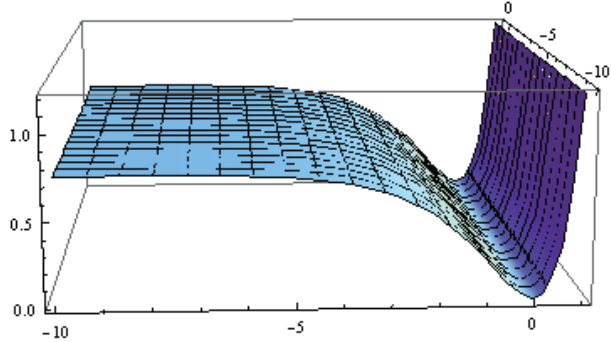}
\caption{The scalar potential $V/M^2$ of the Starobinsky model at $\a=0$ and $\b=0$. It serves as the starting point for
deformations in the moduli space $(\a,\b)$.}
\label{fig:StarobinskyPotential}
\end{center}
\end{figure}

\begin{figure}[ht] 
\begin{center}
\includegraphics[clip,width=15cm]{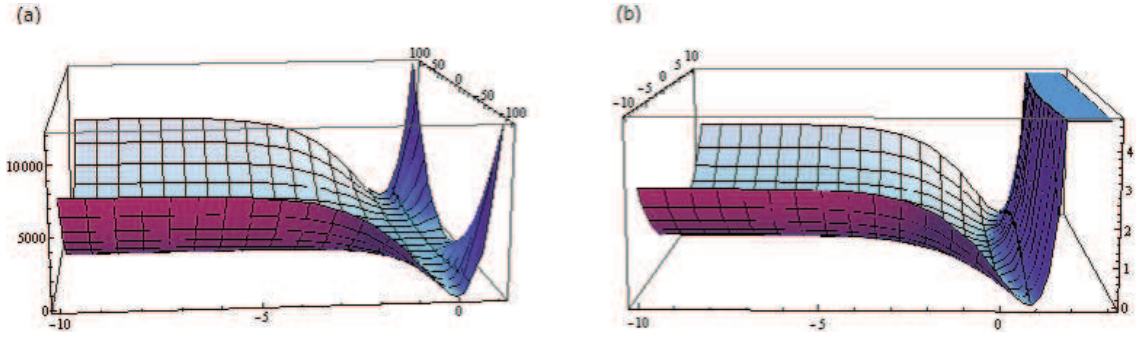}
\caption{(a) The scalar potential $V/M^2$ at $\a=0$ and $\b=1$.\ (b) The scalar potential $V/M^2$ at $\a=0.01$ and $\b=0$.} 
\label{fig:ABPotential1}
\end{center}
\end{figure}

The shape of the Starobinsky scalar potential at $\a=\b=0$ is given in Fig.~\ref{fig:StarobinskyPotential}. In that case the scalar potential does not depend upon $\f$ at all. The one-dimensional deformations of the scalar potential in the $\a$- and $\b$-directions are given by Fig.~\ref{fig:ABPotential1} (a) and (b), respectively.

\begin{figure}[ht] 
\begin{center}
\includegraphics[clip, width=15cm]{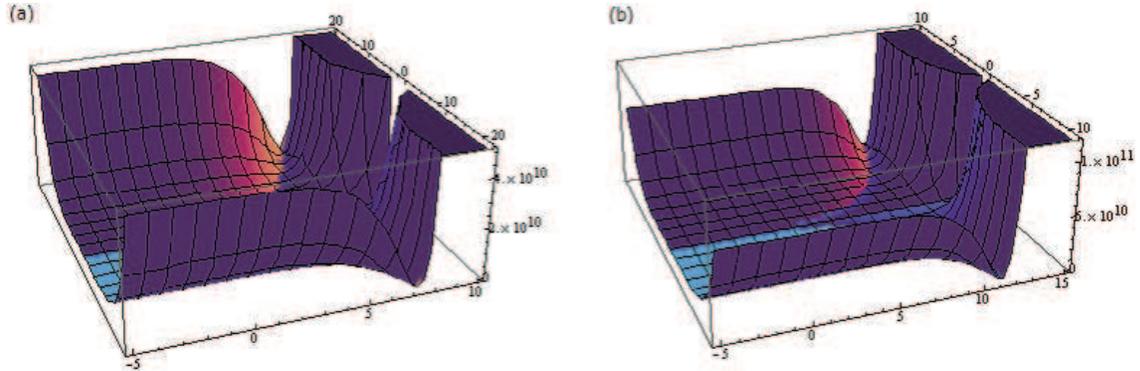}
\caption{(a) The scalar potential $V/M^2$ at $\a=1$ and $\b=1000$.\ (b) The scalar potential $V/M^2$ at $\a=100$ and $\b=10$.} 
\label{fig:ABPotential2}
\end{center}
\end{figure}

The deformed scalar potential in the $\b$-direction (Fig.~\ref{fig:ABPotential1}(a)) at $\a=0$ essentially amounts to rescaling the $M^2$ to the $M^2(1+\b\f^2)$, though the effect of stabilization of $\f$ is already visible in Fig.~\ref{fig:ABPotential1}(a). A change of the parameter  $\b$ at $\a=0$ merely affects the amplitude of CMB fluctuations that fixes the effective scalaron mass, 
$M_{\rm eff}=(3.0 \times10^{-6})(\fracmm{50}{N_e})$. Hence, the Starobinsky inflationary pattern is very robust against changes of $\b$ as long as $\b\f^2\ll 1$ or, simply, when $\beta$ is much less than $1$. The situation does not significantly change under small finite values of the parameter $0<\a\ll 1$, as is illustrated  by Fig.~\ref{fig:ABPotential1}(b). When either $\a$, or $\b$, or both, grow well beyond $1$, the field $\f$ is quickly stabilized, as is illustrated by Fig.~\ref{fig:ABPotential2}, (a) and (b). However, our numerical calculations show that the inflation becomes not viable by failing to get the observed value of the spectral (scalar)  index $n_s$. Therefore, in what follows, we assume that both $\a$ and  $\b$ are well below $1$.

\begin{figure}[ht] 
\begin{center}
\includegraphics[clip, width=8cm]{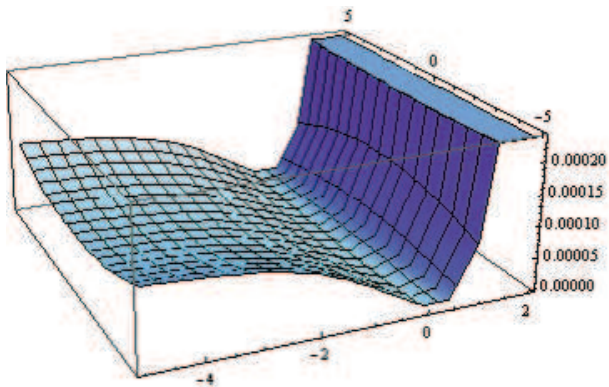}
\caption{The scalar potential $V/M^2$ at $\a=0.01$ and $\b=0.001$.} 
\label{fig:Apotential}
 \end{center}
\end{figure}

As our primary example, we investigate the viable inflationary model specified by the parameters
$\a=0.01$ and $\b=0.001$ in more detail below. The profile of its scalar potential is given in Fig.~\ref{fig:Apotential}.

\begin{figure}[ht] 
\begin{center}
\includegraphics[clip, width=15cm]{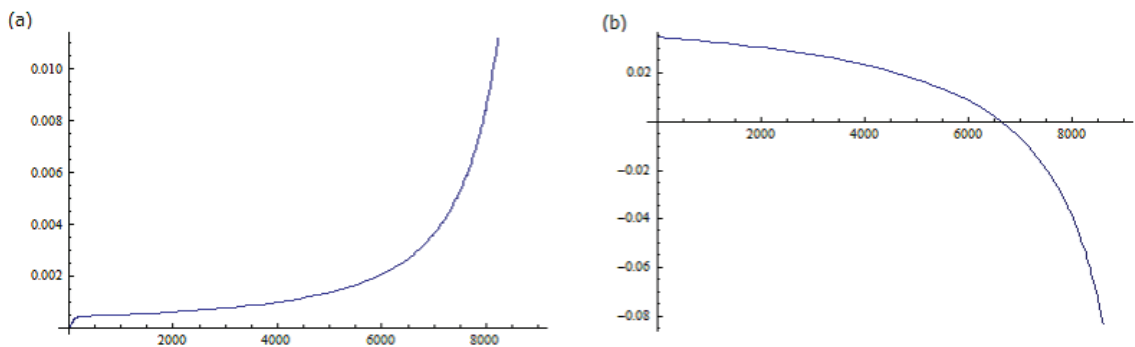}
\caption{(a) The running slow-roll parameter $\e$ at  $\a=0.01$ and $\b=0.001$.\ (b) The running slow-roll parameter $\eta$ at  $\a=0.01$ and $\b=0.001$.} 
\label{fig:runslowroll1}
 \end{center}
\end{figure}

The (time) running of the slow-roll parameters $\e$ and and $\eta$ in our special example is given by 
Fig.~\ref{fig:runslowroll1}(a) and (b), respectively.  The spectral scalar index at the pivot scale is given by $n_s=0.96\pm 0.01$. As to the  tensor-to-scalar ratio $r$, we get $r=0.056\pm 0.003$. The spectral scalar index running
$\a_s \equiv dn_s/d \ln k$ is  $|\a_s| <0.05$  in all our models.

\begin{figure}[ht] 
\begin{center}
\includegraphics[clip, width=15cm]{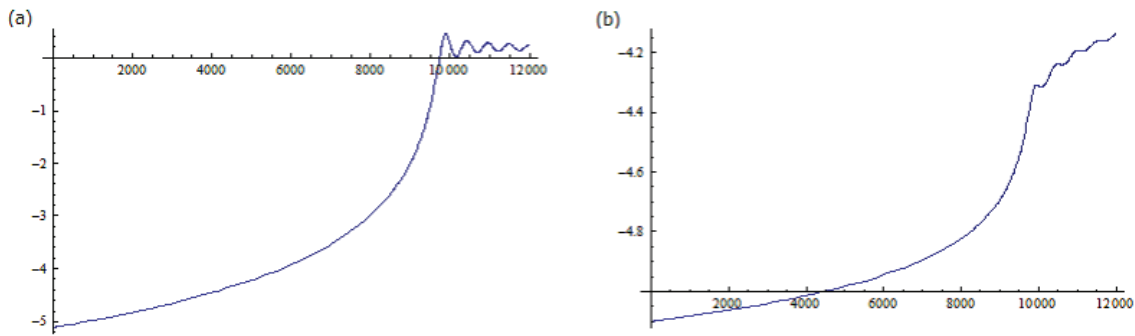}
\caption{(a)The running of $\r$ at $\a=0.01$ and $\b=0.001$.\ (b)The running of $\phi$ at  $\a=0.01$ and $\b=0.001$.} 
\label{fig:runrhophi}
 \end{center}
\end{figure}

To get those results, we used numerical solutions to the background equations of motion, whose graphs  are given
by Fig.~\ref{fig:runrhophi}, (a) and (b), for the fields $\r$ and $\f$, respectively.

Our numerical calculations in this Section support the qualitative conclusion that the Starobinsky inflation is robust against the field dependence in the non-minimal functions $f(\f)$ and  $M(\f)$, as long as the non-minimal coefficients $\a$ and $\b$ are much less than $1$. In other words, the Starobinsky inflation is { \it stable} against small deformations of the non-minimal couplings as long as those deformations are much less than of the order $1$ (in Planck units). In the case of large deformations, inflation persists but is not viable.

\begin{figure}[ht] 
\begin{center}
\includegraphics[clip, width=15cm]{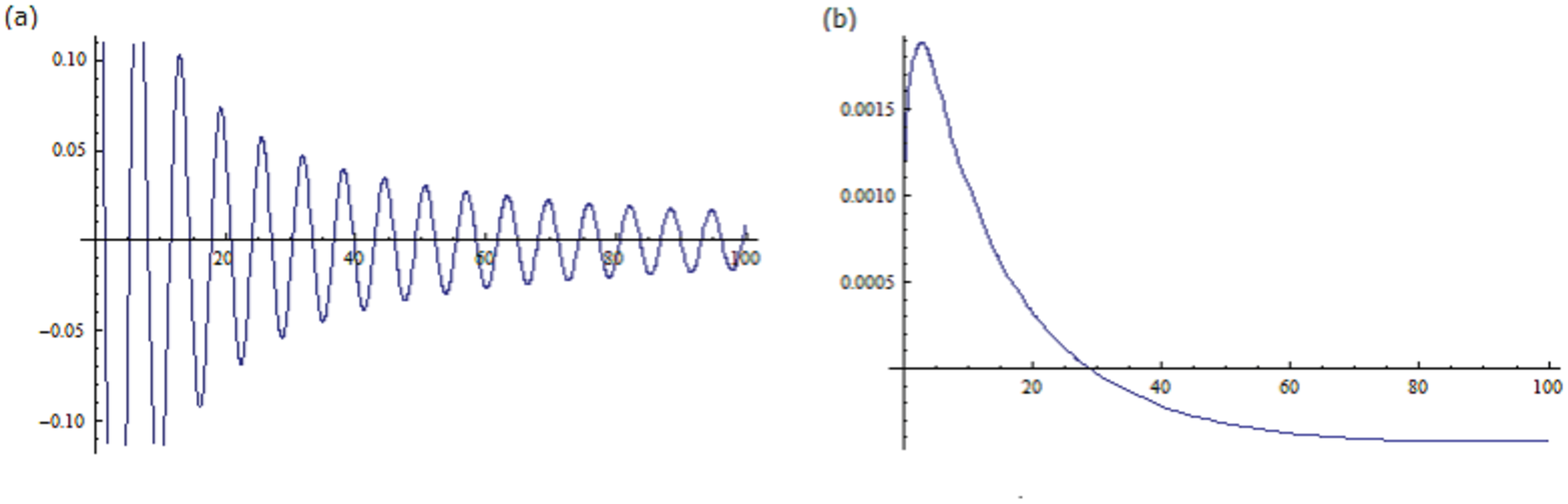}
\caption{(a) The typical behavior of the scalar field $\r$ near its minimum after inflation.\ (b) The typical behavior of the scalar $\f$ near its minimum after inflation.} 
\label{fig:OMScalaronphi}
 \end{center}
\end{figure}

We also found that at the end of inflation the scalaron field $\r$ oscillates near its minimum and thus contributes to (pre)heating, whereas the (matter) scalar field $\f$ does not, approaching a constant value. It can be already
seen in Fig.~\ref{fig:runrhophi}, (a) and (b), but is much better illustrated by our numerical findings in Fig.~\ref{fig:OMScalaronphi}(a) and (b). Actually, the field  $\f$ starts oscillating and thus contributing to the reheating only when the parameter $\a$ is much larger than $1$, however, it does not lead to a viable inflation.

The profile of the $\r$-solution does not significantly change under varying the parameters $\a$ and $\b$. The behavior of the $\f$ scalar during slow-roll and after is more sensitive to the values of the parameters $\a$ and $\b$. In all cases we observe stabilization of $\f$ after slow-roll, as long as one of the parameters $\a$ or $\b$ is positive. It gives another  manifestation of the robustness of the Starobinsky inflation against small changes of the parameters $\a$ and $\b$. 

\begin{figure}[ht] 
\begin{center}
\includegraphics[clip, width=15cm]{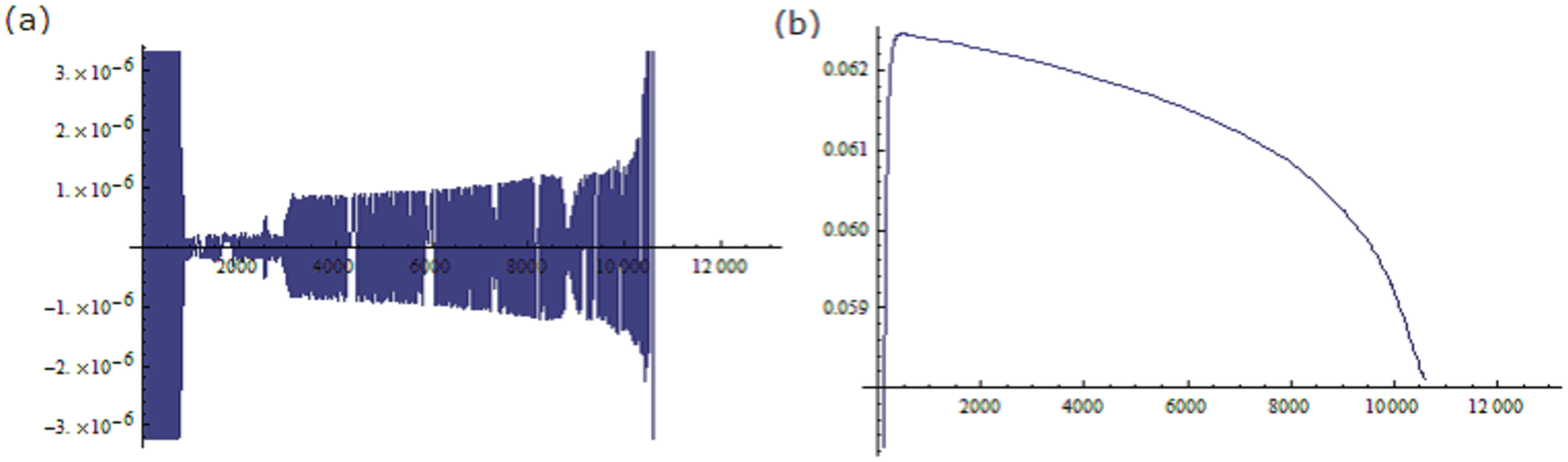}
\caption{(a) The behavior of $\d\r$. (b) The behavior of $\d\f$.} 
\label{fig:runfluctuations}
 \end{center}
\end{figure}

Finally, the numerical solutions to the perturbation equations for fluctuations $\d\r$ and $\d\f$ on the background specified
by Fig.~\ref{fig:runrhophi}, in our primary example with the parameters $\a=0.01$ and $\b=0.001$, are presented in
Fig.~\ref{fig:runfluctuations}.

\section{Conclusion}\label{sec:Con}

We found that the Starobinsky inflation is robust against mixing scalaron with another (matter) scalar via non-minimal interactions of the latter with both $R$ and $R^2$ terms in the original (Jordan) frame, as long as the non-minimal field couplings are much smaller than one (in the Planck units). The non-minimal couplings were introduced by promoting the parameters of the original Starobinsky model to the (matter scalar) field-dependent functions, under the additional
restriction of renormalizability of matter in the classical gravitational background.

We confirmed numerically that the inflationary trajectory in our two-field inflationary models remains close to the single-field attractor solution in the original Starobinsky model \cite{star1} under adding small non-minimal couplings to the $R$ and $R^2$ terms in Eq.~(\ref{ourm1}). Our main statement is reflected in the title of our paper by calling our two-field inflationary models   the {\it Starobinsky-like} ones. Though our numerical solutions to the dynamical equations (Sec.~5) were obtained by using some initial conditions for inflation, we found that the dependence of our solutions upon small changes in the initial conditions is weak and rather unimportant.  It is related to the facts that (i) our numerical solutions also exhibit an attractor-type behavior
(see e.g., Refs.~\cite{klin1,klin2} for more),  and (ii) our scalar potentials do not have ridges that are generically present 
in multi-field inflation caused by non-minimal couplings and whose presence leads to strong dependence upon the initial conditions at the onset of inflation \cite{d1}.

The two-field Starobinsky-like inflation becomes not viable when any of the non-minimal parameters is of the order one or larger. Our results are complementary to the findings of Ref.~\cite{uk2} where the robustness of the Starobinsky inflation was established in another two-field Starobinsky-like limit with $\a=\b=0$ and a non-vanishing mass term of the matter scalar.

The main difference of our two-field inflationary models against the single-field Starobinsky model is the presence of isocurvature perturbations. However, those perturbations turn out to be very small and (currently) undetectable. As was argued in Ref.~\cite{dsprl}, significant isocurvature perturbations in generic multi-field inflationary models with 
non-minimal couplings may account for the observed low power in the CMB angular power spectrum of temperature anisotropies at low multipoles \cite{planck13}. However, in our models the isocurvature perturbations are not amplified enough to be the reason for that observation.

Though we did not investigate primordial non-Gaussianities in our Starobinsky-like two-field inflationary models, we
expect them to be negligible, like the original (single-field) Starobinsky model. 

The field-dependent couplings are quite natural from the viewpoint of {\it string theory} where all coupling constants
are given by expectation values of scalar fields. As regards the physical meaning of our two scalars from the viewpoint of string theory, it is conceivable that scalaron is related to string theory dilaton, whereas another (matter) scalar is given by one of the moduli arising from superstring compactification. A detailed investigation of the possible connection of our models to string  theory is beyond the scope of this paper.

\section*{Acknowledgements}

This work was supported by a Grant-in-Aid of the Japanese Society for Promotion of Science (JSPS) under No.~26400252, 
the World Premier International Research Center Initiative (WPI Initiative), MEXT, Japan, the special TMU Fund for International Research, and the Competitiveness Enhancement Program of the Tomsk Polytechnic University in Russia.
The authors are grateful to D. Kaiser and S. Vagnozzi for discussions and correspondence, and to the referee for his
careful reading of our submission and critical remarks.

\vglue.2in

\section*{Appendix A: cosmological perturbations, their power spectra and evolution at the horizon 
and super-Hubble scales}

\renewcommand{\theequation}{A.\arabic{equation}}
\setcounter{equation}{0}

\vglue.2in
{\it Linear perturbations:}
\vglue.1in

The standard form of scalar-perturbed space-time metric in the longitudinal gauge (when the off-diagonal spatial components of the stress-energy tensor vanish) is given by
\be
ds^2 = (1+ 2\F)dt^2 - a^2(1-2\F)d\mathbf{x}^2.
\ee 

One can decompose the scalar fields into their backgrounds and perturbations as follows:
\be
\r (t, \mathbf{x}) = \r(t) + \d\r (t, \mathbf{x}),
\ee
\be
\f (t, \mathbf{x}) = \f(t) + \d\f (t, \mathbf{x}).
\label{eq13}
\ee

The Fourier components of the perturbations are denoted by $\d\r_{\mathbf{k}}(t)$ and $\d\f_{\mathbf{k}}(t)$, respectively.  When omitting the subscript ${\bf k}$  for simplicity, as is common in the literature,  the perturbed Klein-Gordon equations of motion read
\be
\begin{split}
&\ddot{\d\r} + 3H\dot{\d\r} + \left[ \fr{k^2}{a^2} +V_{,\r\r} - (b_{,\r\r} + 2b_{,\r}^2)\dot{\f}^2 \ex^{2b}\right] \d\r +V_{,\r\f}\d\f -2b_{,\r}\ex^{2b}\dot{\f}\dot{\d\f} \\
&= 4\dot{\r}\dot{\F} -2V_{,\r}\F~, \\
\end{split}
\label{perturbed1}
\ee
\be
\begin{split}
&\ddot{\d\f} + (3H+2b_{,\r})\dot{\d\f} +\left[\fr{k^2}{a^2} +\ex^{-2b}V_{,\f\f}\right]\d\f + 2b_{,\r}\dot{\f}\dot{\d\r} +\left[ \ex^{-2b}(
V_{,\f\r}-2b_{,\r}V_{,\f}) +2b_{,\r\r}\dot{\r}\dot{\f}\right]\d\r \\
&= 4\dot{\f}\dot{\F} - 2\ex^{-2b}V_{,\f}\F~.
\end{split}
\label{perturbed2}
\ee

The Einstein equations lead to the energy and momentum constraints as follows:
\be
3H\left( \dot{\F} + H\F \right) +\dot{H}\F +\fr{k^2}{a^2}\F = -\fr{1}{2} \left( \dot{\r}\dot{\d\r} + \ex^{2b}\dot{\f}\dot{\d\f} 
+b_{,\f}\ex^{2b}\dot{\f}^2\d\r + V_{,\r}\d\r + V_{,\f}\d\f \right)~,
\label{constraint1}
\ee
\be
\dot{\F} + H\f = \fr{1}{2} \left(\dot{\r}\d\r + \ex^{2b}\dot{\f}\d\f \right)~.
\label{constraint2}
\ee

In terms of the {\it Mukhanov-Sasaki} (MS) variables \cite{mukh}
\be
Q_{\r} \equiv \d\r + \fr{\dot{\r}}{H}\F \ \ \ {\rm and}\ \ \ Q_{\f} \equiv \d\f + \fr{\dot{\f}}{H}\F
\ee
the perturbed Klein-Gordon equations take the form
\be
\ddot{Q_{\r}} +3H\dot{Q_{\r}} -2\ex^{2b}b_{,\r}\dot{\f}\dot{Q_{\f}} + \left( \fr{k^2}{a^2} + C_{\r\r}\right) Q_{\r} + C_{\r\f}Q_{\f} =0~,
\label{eq18}
\ee
\be
\ddot{Q_{\f}} +3H\dot{Q_{\f}} + 2b_{,\r}\dot{\r}\dot{Q_{\f}} +2b_{,\r}\dot{\f}\dot{Q_{\r}} + \left(\fr{k^2}{a^2} + C_{\f\f} \right)Q_{\f} + C_{\f\r}Q_{\r} =0~,
\label{eq19}
\ee
where the background equations and the energy-momentum constraints above have been used in the notation \cite{lal},
\be
C_{\r\r} = -2\ex^{2b}b_{,\r}^2\dot{\f}^2 +3\dot{\r}^2 -\fr{\ex^{2b}\dot{\r}^2\dot{\f}^2}{2H^2} -\fr{\dot{\r}^4}{2H^2} 
-\ex^{2b}b_{,\r\r}\dot{\f}^2 +\fr{2V_{,\r}}{H} +V_{,\r\r}~~,
\label{eq20}
\ee
\be
C_{\r\f} = 3\ex^{2b}\dot{\r}\dot{\f} -\fr{\ex^{4b}\dot{\r}\dot{\f}^3}{2H^2} -\fr{\ex^{2b}\dot{\r}^3\dot{\f}}{2H^2} +\fr{\dot{\r}V_{,\f}}{H} 
+\fr{\ex^{2b}\dot{\f}V_{,\r}}{H} +V_{,\r\f}~~,
\label{eq21}
\ee
\be
C_{\f\f} = 3\ex^{2b}\dot{\f}^2 -\fr{\ex^{4b}\dot{\f}^4}{2H^2} -\fr{\ex^{2b}\dot{\r}^2\dot{\f}^2}{2H^2} +\fr{2\dot{\f}V_{,\f}}{H} 
+\ex^{-2b}V_{,\f\f}~~,
\label{eq22}
\ee
\be
C_{\f\r} = 3\dot{\r}\dot{\f} -\fr{\ex^{2b}\dot{\r}\dot{\f}^3}{2H^2} -\fr{\dot{\r}^3\dot{\f}}{2H^2} +2b_{,\r\r}\dot{\r}\dot{\f}
 -2\ex^{-2b}b_{,\r}V_{,\f} +\fr{\ex^{-2b}\dot{\r}V_{,\f}}{H} +\fr{\dot{\f}V_{,\r}}{H} +\ex^{-2b}V_{,\r\f}~.
\label{eq23}
\ee

\vglue.2in
{\it Adiabatic and entropy perturbations:}
\vglue.1in

It is common in the literature to decompose cosmological linear scalar perturbations into two directions that are either parallel or orthogonal to the trajectory in the field space \cite{mukh}. The first type of perturbations is called {\it curvature} (or {\it adiabatic}) perturbations, whereas the second type is called {\it isocurvature} (or {\it entropy}) perturbations, respectively. It can be done by introducing the linear combinations (they do not refer to new scalar fields)
\be
\d\s \equiv \cos\q\d\r + \sin\q\ex^{b}\d\f   \ \ \ \ \ {\rm and} \ \ \ \ \ \d s \equiv -\sin\q\d\r +\cos\q\ex^{b}\d\f,
\label{eq24}
\ee
where we have used the notation
\be
\cos\q \equiv \fr{\dot{\r}}{\dot{\s}}~~,\quad \sin\q \equiv \fr{\dot{\f}\ex^{b}}{\dot{\s}}~~,\quad
 \dot{\s} \equiv \sqrt{\dot{\r}^2 + \ex^{2b}\dot{\f}^2}~~.
\label{eq25}
\ee

The corresponding MS variables $ Q_{\s} \equiv \d\s + \fr{\dot{\s}}{H}\F$
are given by the linear combinations 
\be
Q_{\s} \equiv \cos\q \ Q_{\r} + \sin\q \ \ex^{b}Q_{\f}~,\qquad \d s \equiv -\sin\q\ Q_{\r} + \cos\q \ \ex^{b}Q_{\f}~.
\label{eq27}
\ee
 
The gauge-invariant quantity, known as the  {\it co-moving} curvature perturbation \cite{mukh},  reads in terms of $Q_{\s}$ as
follows:
\be
\mathcal{R} \equiv \fr{H}{\dot{\s}}Q_{\s}~,
\label{eq28}
\ee
while the renormalized entropy (iso-curvature) perturbation \cite{mukh} is given by
\be
\mathcal{S} \equiv \fr{H}{\dot{\s}}\d s~.
\label{eq29}
\ee

In terms of the adiabatic and entropy "vectors" in field space, defined by \cite{lal} 
\be
E_{\s}^I = (\cos\q,\  \ex^{-b}\sin\q),\ \ \ \ \ E_{s}^{I} = (-\sin\q,\  \ex^{-b}\cos\q),\quad  I= \left\{ \r,\ \f \right\}~,
\label{eq30}
\ee
the corresponding first order derivatives are
\be
V_{\s} = E_{\s}^I V_{,I},\qquad V_{s} = E_{s}^I V_{,I}~,
\label{eq31}
\ee
and the second order derivatives are
\be
V_{\s\s} = E_{\s}^I E_{\s}^J V_{,IJ}, \quad V_{\s s} = E_{\s}^I E_{s}^J V_{,IJ}, \quad V_{ss} = E_s^I E_s^J V_{,IJ}~.
\label{eq32}
\ee

Given the notation above, the background equations of motion in the adiabatic and entropy directions are
\be
\ddot{\s} + 3H\dot{\s} +V_{\s} = 0~,
\label{eq33}
\ee
\be
\dot{\q} = -\fr{V_{s}}{\dot{\s}} -b_{,\r}\dot{\s}\sin\q~,
\label{eq34}
\ee
respectively. The equations of motion for perturbations are given by
\be
\ddot{Q_{\s}} +3H\dot{Q_{\s}} +\left(\fr{k^2}{a^2} + C_{\s\s}\right)Q_{\s} +\fr{2V_s}{\dot{\s}}\dot{\d s} +C_{\s s}\d s =0~,
\label{EOM for pb1}
\ee
\be
\ddot{\d s} +3H\dot{\d s} +\left( \fr{k^2}{a^2} +C_{ss}\right)\d s -\fr{2V_s}{\dot{\s}}\dot{Q_{\s}} +C_{s\s}Q_{\s} =0~,
\label{EOM for pb2} 
\ee
where we have used the notation \cite{lal} again:
\be
C_{\s\s} = V_{\s\s} -\left(\fr{V_s}{\dot{\s}}\right)^2 +\fr{2\s V_{\s}}{H} +3\dot{\s}^2 -\fr{\dot{\s}^4}{2H^2} -b_{,\r}\left(s_{\q}^2 c_{\q} V_{\s} +(c_{\q}^2 +1)s_{\q}V_s \right)~,
\label{eq37}
\ee
\be
C_{\s s} = \fr{6HV_s}{\dot{\s}} +\fr{2V_{\s}V_{s}}{\dot{\s}^2} +2V_{\s s} +\fr{\dot{\s}V_s}{H} + 2b_{\r}\left(s_{\q}^3V_{\s} -c_{\q}^3V_s \right)~,
\label{eq38}
\ee
\be
C_{ss} = V_{ss} -\left(\fr{V_s}{\dot{\s}}\right)^2 +b_{,\r}(1+s_{\q}^2)c_{\q}V_{\s} +b_{,\r}c_{\q}^2s_{\q}V_s -\dot{\s}^2(b_{,\r\r}+b_{,\r}^2)~,
\label{eq39}
\ee
\be
C_{s\s} = -\fr{6HV_s}{\dot{\s}} -\fr{2V_{\s}V_{s}}{\dot{\s}^2} +\fr{\dot{\s}V_{s}}{H}~,
\label{eq40}
\ee
with the $s_{\q}$ and $c_{\q}$ standing for the $\sin\q$ and $\cos\q$, respectively.

\vglue.2in
{\it Perturbation spectra:}
\vglue.1in

We are now in a position to study the perturbation spectra of the two-field inflation and evolution of its perturbations. The power spectra of the adiabatic and entropy perturbations are given by the correlation functions \cite{lal}
\be
\left< Q_{\s_{\mathbf{k}}}^{*} Q_{\s_{\mathbf{k}^{\prime}}} \right> = \fr{2\p^2}{k^3}\cp_{Q_{\s}}(k)\d(\mathbf{k} - \mathbf{k}^{\prime})~,
\ee
\be
\left< \d s_{\mathbf{k}}^{*} \d s_{\mathbf{k}^{\prime}} \right> = \fr{2\p^2}{k^3}\cp_{\d s}(k)\d(\mathbf{k} - \mathbf{k}^{\prime})~,
\ee
\be
\left< Q_{\s_{\mathbf{k}}}^{*} \d s_{\mathbf{k}^{\prime}} \right> = \fr{2\p^2}{k^3}\cc_{Q_{\s}\d s}(k)\d(\mathbf{k} - \mathbf{k}^{\prime})~.
\ee  

The cosmologically important  scales are given by (i) the horizon crossing (inside the Hubble scale) and (ii) the scales over the Hubble scale, so that it is natural to evaluate the correlation functions at those scales, along the standard procedure \cite{lal}.

\vglue.2in
{\it Evolution of perturbations at the horizon crossing:}
\vglue.1in

In terms of the conformal time $\t = \int \fr{1}{a(t)} dt$ and the new variables
\be
u_{\s} = aQ_{\s}~,\qquad u_s = a\d s~,
\ee
equations (\ref{EOM for pb1}) and (\ref{EOM for pb2}) can be rewritten to
\be
u_{\s}^{\prime\prime} +\fr{2V_s}{\dot{\s}}au_{s}^{\prime} +\left[k^2 -\fr{a^{\prime\prime}}{a} +a^2C_{\s\s}\right] u_{\s} +\left[ -\fr{2V_s}{\dot{\s}}a^{\prime} +a^2C_{\s s}\right]u_s =0~,
\ee
\be
u_{s}^{\prime\prime} -\fr{2V_s}{\dot{\s}}au_{\s}^{\prime} +\left[k^2 -\fr{a^{\prime\prime}}{a} +a^2C_{ss}\right] u_s +\left[\fr{2V_s}{\dot{\s}}a^{\prime} +a^2C_{s\s}\right]u_{\s} =0~,
\ee
where the primes denote the derivative with respect to the conformal time $\t$.

In the slow-roll approximation these equations can be further simplified as 
\be
\left[\left( \fr{d^2}{d\t^2} + k^2 -\fr{2+3\e}{\t^2}\right)\mathbf{1} +2\mathbf{E}\fr{1}{\t}\fr{d}{d\t} +\mathbf{M}\fr{1}{\t^2} \right]
\left(
\begin{array}{c}
u_{\s} \\
u_{s} \\
\end{array}
\right)
=0~,
\label{simplified}
\ee
with the notation 
\be
\mathbf{E} =
\left(
\begin{array}{cc}
0 & -\h_{\s s} \\
\h_{\s s} & 0 \\
\end{array}
\right)
+
\left(
\begin{array}{cc}
0 & \x s_{\q}^3 \\
-\x s_{\q}^3 & 0 \\
\end{array}
\right),
\label{E}
\ee
\be
\mathbf{M} =
\left(
\begin{array}{cc}
-6\e + 3\h_{\s\s} & 4\h_{\s s} \\
2\h_{\s s} & 3\h_{ss} \\
\end{array}
\right)
+
\left(
\begin{array}{cc}
3\x s_{\q}^2 c_{\q} & -4\x s_{\q}^3 \\
-2\x s_{\q}^3 & -3\x c_{\q}(1+s_{\q}^2) \\
\end{array}
\right),
\label{M}
\ee
and
\be
\x \equiv \sqrt{2}b_{,\r}\sqrt{\e}~.
\ee

In Eqs.~(\ref{E}) and (\ref{M}) we kept only the linear terms with respect to $b_{,\r}$ because  
it is suppressed by the $M_{\rm Pl}$ and $b_{,\r\r}=0$ in our case (\ref{mod}). The terms proportional to $\x$ are
written down separately, in order to emphasize the difference between the canonical and non-canonical kinetic terms. According to Ref.~\cite{lal}, it is convenient to introduce $\mathbf{L}$ and $\mathbf{Q}$ as
\be
2\mathbf{L} =\fr{2\mathbf{E}}{\t}~,\quad \mathbf{Q}=\left(k^2-\fr{2+3\e}{\t^2}\right)\mathbf{1} +\fr{\mathbf{M}}{\t^2}~,
\ee
and then rewrite Eq.~(\ref{simplified}) to the standard (in mathematical physics) form
\be
u^{\prime\prime} + 2\mathbf{L}u^{\prime} +\mathbf{Q}u =0~.
\label{u L Q}
\ee

As the next step, again following Ref.~\cite{lal}, let us introduce the time-dependent matrix $\mathbf{R}$ which satisfies $\mathbf{R}^{\prime} = -\mathbf{LR}$, and the new vector $v$ defined by $u=\mathbf{R}v$. Then the equation above can be resolved for $v$ as
\be
v^{\prime\prime} + \mathbf{R}^{-1}\left(-\mathbf{L}^2 -\mathbf{L}^{\prime} + \mathbf{Q} \right) \mathbf{R}v =0~,
\label{eq in v}
\ee
where
\be
-\mathbf{L}^2-\mathbf{L}^{\prime} \simeq \frac{1}{\t^2}\mathbf{E}
\ee
in the linear order with respect to the slow-roll parameters. It follows
\be
-\mathbf{L}^2 -\mathbf{L}^{\prime} +\mathbf{Q}  \simeq \left(k^2 - \frac{2+3\e}{\t^2}\right)\mathbf{1} + \frac{1}{\t^2}\left(\mathbf{E}+\mathbf{M}\right)~,
\ee 
where the second term reads
\be
\fr{1}{\t^2}\left(\mathbf{E}+\mathbf{M}\right) = \fr{3}{\t^2}
\left(
\begin{array}{cc}
-2\e + \h_{\s\s} +\x s_{\q}^2c_{\q}  &  \h_{\s s} - \x s_{\q}^3 \\
\h_{\s s} - \x s_{\q}^3  &  \h_{ss} -\x c_{\q}(1+s_{\q}^2)
\end{array}
\right)~.
\label{E+M}
\ee

It is usually assumed in the literature that the slow-roll parameters vary slowly enough during the few e-folds when the inflationary scale crosses the Hubble radius. In that case, one can replace the time-dependent matrix with its value at the Hubble crossing. In other words, the matrix on the r.h.s. of Eq.~(\ref{E+M}) is supposed to be evaluated at $k=aH$, so that the remaining time dependence only exist in the overall coefficient $3/{\t^2}$. Then one can always diagonalize this matrix by using a time-independent rotation matrix
\be
\mathbf{\tilde{R}_{*}} =
\left(
\begin{array}{cc}
\cos{\Q_{*}} & -\sin{\Q_{*}} \\
\sin{\Q_{*}} & \cos{\Q_{*}} \\
\end{array}
\right)~,
\ee
so that
\be
\mathbf{\tilde{R}_{*}}^{-1} \left(\mathbf{E} +\mathbf{M} \right) \mathbf{\tilde{R}_{*}} = \mathrm{Diag} \left( \tilde{\l}_1, \tilde{\l}_2 \right)~,
\ee
where the star subscript  refers to the horizon crossing. 

Since $\mathbf{R}$ varies slowly around the Hubble crossing, one can replace $\mathbf{R}$ by $\mathbf{R}_{*}$ .
 When using the notation \cite{lal}
\be
\tilde{\l}_1 + \tilde{\l}_2 = 3\left(\h_{\s\s} +\h_{ss} -2\e -\x c_{\q}\right)~,
\label{l12-1}
\ee
\be
( \tilde{\l}_1 - \tilde{\l}_2 ) \sin{2\Q_{*}} = 6 \left(\h_{\s s} -\x s_{\q}^3 \right)~,
\label{l12-2}
\ee
\be
( \tilde{\l}_1 - \tilde{\l}_2 ) \cos{2\Q_{*}} = 3 \left( \h_{\s\s} - \h_{ss} -2\e +\x c_{\q}(1+2s_{\q}^2) \right)~,
\label{l12-3}
\ee
at $k=aH$, and
\be
w = \mathbf{\tilde{R}}_{*}^{-1} \mathbf{R}_{*} v,
\ee
and one can also rewrite Eq.~(\ref{u L Q}) to
\be
w_{A}^{\prime\prime} + \left[ k^2 -\fr{1}{\t^2}(2+3\l_{A}) \right] w_{A} =0 \ \ \ \ \ (A=1,2),
\label{eq in w}
\ee
with
\be
\l_{A} = \e - \fr{1}{3}\tilde{\l}_{A}~.
\ee

The solution to Eq.~(\ref{eq in w}) with the proper asymptotic behavior reads \cite{lal}
\be
w_{A} = \fr{\sqrt{\p}}{2}\mathrm{exp}\left(\fr{i(\m_{A}+\fr{1}{2})\p}{2}\right) \sqrt{-\t}H^{(1)}_{\m_{A}}(-k\t)e_{A}(k)
\ee
in terms of the Hankel function $H^{(1)}_{\m}$  of the first kind and of the order $\m_A$, where \cite{lal}
\be
\m_{A} = \sqrt{\fr{9}{4} + 3\l_{A}}~,
\ee
and $e_{A}$, $A=1,2$, are the independent orthonormal (Gaussian) random variables,
\be
\left< e_{A}(k) \right> =0~,\quad \left< e_{A}(k)e^{*}_{B}(k^{\prime}) \right> = \d_{AB}\d^{(3)}(k-k^{\prime})~.
\ee

Because of independence of $w_1$ and $w_2$, the correlations of $u_{\s}$ and $u_{s}$ around the Hubble crossing can be expressed in terms of $Q_{\s}$ and $\d s$ as
\be
a^2 \left< Q_{\s}^{\dagger} Q_{\s} \right> = \cos^2{\Q_{*}} \left<w_1^{\dagger} w_1\right> + \sin^2{\Q_{*}} \left<w_2^{\dagger} w_2 \right>~,
\ee
\be
a^2 \left<\d s^{\dagger} \d s\right> = \sin^2{\Q_{*}} \left< w_1^{\dagger} w_1 \right> + \cos^2{\Q_{*}} \left< w_2^{\dagger} w_2 \right>~,
\ee
\be
a^2 \left<\d s^{\dagger} Q_{\s} \right> = \fr{1}{2} \sin{2\Q_{*}}\left( \left< w_1^{\dagger} w_1 \right> - \left< w_2^{\dagger} w_2 \right> \right)~,
\ee
where we have
\be
\left< w_{A}^{\dagger} w_{A} \right> = \fr{\p}{4} (-\t) \left| H^{(1)}_{\m_{A}} (-k\t) \right|^2 \equiv \fr{1}{2k} \fr{1}{(k\t)^2} \cg_{A} (-k\t)~.
\ee
Therefore, after taking into account that 
\be
a \simeq -\fr{1+\e_{*}}{H_{*}\t}~,
\ee
one finds \cite{lal}
\be
\cp_{Q_{\s}} = \left(\fr{H_{*}}{2\p}\right)^2 (1-2\e_{*}) \left[ \cos^2{\Q_{*}} \cg_{1}(-k\t) + \sin^2{\Q_{*}} \cg_{2}(-k\t) \right],
\ee
\be
\cp_{\d s} = \left(\fr{H_{*}}{2\p}\right)^2 (1-2\e_{*}) \left[ \sin^2{\Q_{*}} \cg_{1}(-k\t) + \cos^2{\Q_{*}} \cg_{2}(-k\t) 
\right]~,
\ee
\be
\cc_{Q_{\s}\d s} =  \left(\fr{H_{*}}{2\p}\right)^2 (1-2\e_{*}) \fr{\sin{2\Q_{*}}}{2} \left[\cg_{1}(-k\t) - \cg_{2}(-k\t)\right]~.
\ee

Given $\l_A \ll 1$, \rm{i.e.} $\m_A \simeq \fr{3}{2} + \l_A$, one can further simplify the result above, by expanding
$\cg_{A}(x)$ as follows:
\be
\cg_{A}(x) = \fr{\p}{2} x^3 \left| H_{\fr{3}{2}}(x) \right|^2 (1 + 2\l_{A} g(x) ) = (1+x^2) (1 + 2\l_A g(x))~,
\ee 
where the new function has been introduced,
\be
g(x) = \mathrm{Re} \left( \fr{1}{H_{\fr{3}{2}}^{(1)}(x)} \left. \fr{dH_{\m}^{(1)}(x)}{d\m} \right|_{\m=\fr{3}{2}} \right)~.
\ee

It follows that the power spectra and the correlations of curvature and entropy perturbations are
\be
\cp_{\car} = \left( \fr{H^2_{*}}{2\p\dot{\s}_{*}} \right) (1+k^2\t^2) \left[ 1+2\e_{*}+(6\e_{*} -2\h_{\s\s *} -2 \x_{*} s_{\q}^2 c_{\q}) g\left(\fr{k}{aH_{*}} \right)\right]~,
\label{ps si}
\ee
\be
\cp_{\cs} = \left( \fr{H^2_{*}}{2\p\dot{\s}_{*}} \right)^2 (1+k^2\t^2) \left[ 1-2\e_{*} + (2\e_{*} -2\h_{ss*} +2\x_{*}(1+s^2_{\q *})c_{\q *}) g\left(\fr{k}{aH_{*}} \right)\right]~,
\label{ps s}
\ee
\be
\cc_{\car\cs} = \left( \fr{H^2_{*}}{2\p\dot{\s}_{*}} \right)^2 (1+k^2\t^2) (2\x_{*} s_{\q}^3 -2 \h_{\s s*}) g\left(\fr{k}{aH_{*}} \right)~.
\label{cor}
\ee

\vglue.2in
{\it Evolution of perturbations on super-Hubble scales:}
\vglue.1in

A two-field inflationary model is reduced to a single field inflationary model when the isocurvature 
perturbations are suppressed. Then the adiabatic spectrum takes the form
\be
\cp_{\car}^{SH}(k) \simeq \fr{H^4}{4\p \dot{\s}}~.
\label{single}
\ee
However, it does not apply to our model (Sec.~2) in a generic case where it does not reduce to the
Higgs or the Starobinsky (single field) inflationary model.

The existence of the isocurvature modes is a {\it generic} feature of two-field inflationary models, and it is going to affect adiabatic perturbations also during the super-Hubble scale evolution, so that Eq.~(\ref{single}) does not apply, in general. To get the power spectra and the correlation functions in that case, one should solve the coupled system of Eqs.~(\ref{EOM for pb1}) and (\ref{EOM for pb2}). A numerical approach is the only way in most cases. 

However, in some spacial cases, when the slow-roll approximation is at work, one may analytically solve the equations of motion on the super-Hubble scales too.  The example considered in Ref.~\cite{lal} was Eqs.~(\ref{EOM for pb1}) and (\ref{EOM for pb2}) in the slow roll approximation, 
\be
\dot{Q_{\s}} \simeq AHQ_{\s} +BH\d s \quad {\rm and} \quad \dot{\d_s} \simeq DH\d_s~,
\label{SH ABD}
\ee  
where 
\be
A = -\h_{\s\s} +2\e -\x c_{\q} s_{\q}^2~,
\ee
\be
B = -2\h_{\s s} + 2\x s_{\q}^3 \simeq 2\fr{d\q}{dN} -2\x s_{\q}~,
\ee
\be
D = -\h_{ss} + \x c_{\q}(1 + s_{\q}^2)~.
\ee

Equation (\ref{SH ABD}) implies that adiabatic and isocurvature perturbations have strong interaction unless the isocurvature perturbations rapidly decay. For {\it constant} values of  $(A, B, D)$, Eq.~(\ref{SH ABD})  can be solved as \cite{lal}
\be
Q_{\s}(N) \simeq \ex^{AN}Q_{\s *} +\fr{B}{D-A} \left( \ex^{DN} -\ex^{AN} \right) \d s_{*}~,
\ee
\be
\d s(N) \simeq \ex^{DN} \d s_{*}~,
\ee
where the number $N$ of the e-folds after the Hubble crossing has been introduced.

Given 
\be
\left( \fr{H}{\dot{\s}} \right) \simeq \left( \fr{H_{*}}{\dot{\s_{*}}} \right) \ex^{-AN}~,
\ee
one can easily find the power spectra and the correlation functions as \cite{lal}
\be
\cp^{SH}_{\car}(N) \simeq \bar{\cp}_{\car_{*}} + \bar{\cp}_{\cs_{*}} \left( \fr{B}{\g} \right)^2 \left( \ex^{\g N} -1\right)^2 + 2 \bar{\cc}_{\car\cs *}\fr{B}{\g} \left( \ex^{\g N} -1 \right)~,
\label{eq89}
\ee
\be
\cp^{SH}_{\cs}(N) \simeq \bar{\cp}_{\cs *}\ex^{2\g N}~,
\label{eq90}
\ee
\be
\cc^{SH}_{\car\cs}(N) \simeq \bar{\cc}_{\car\cs *} \ex^{\g N} + \bar{\cp}_{\cs *} \fr{B}{\g} \ex^{\g N} \left( \ex^{\g N} -1 \right)~,
\label{eq91}
\ee
where $\g = D-A $, and the $\bar{\cp}_{\car *},\ \bar{\cp}_{\cs *} $ and $\ \bar{\cc}_{\car\cs *}$ are supposed to be evaluated in the asymptotic limits of Eqs.~(\ref{ps si}),  (\ref{ps s})   and   (\ref{cor}), respectively, \rm{i.e.}\  at $k\t \rightarrow 0$.

Unfortunately, as was already noticed in Ref.~\cite{lal}, the {\it constant} slow-roll approximation does not hold for many e-folds, and breaks down long before the exit from inflation. In another analytically treatable case, with the mass terms as the scalar potential and
 the canonical kinetic terms for scalars, the curvature and iso-curvature perturbations were computed in Ref.~\cite{sp}. 
 
\vglue.2in


\begin{thebibliography}{99}
\bibitem{star1}   A.~A.~Starobinsky,  Phys.\ Lett.\ B {\bf 91} (1980) 99.
\bibitem{star2} A.~A. Starobinsky,  {\it Nonsingular model of the Universe with the quantum 
gravitational de Sitter stage and its observational consequences}, in the Proceedings of the 
2nd Intern. Seminar on “Quantum Theory of Gravity”, Moscow, 13–15 October 1981 (INR Press,
Moscow, 1982), p. 58; reprinted in "Quantum Gravity", eds. M.~A. Markov and P.~C. West 
(Plenum Publ. Co., New York, 1984), p. 103.
\bibitem{mchi}  V.~F. Mukhanov and G.~V. Chibisov, JETP Lett. {\bf 33} (1981) 532.
\bibitem{star3}   A.~A.~Starobinsky, Sov. Astron. Lett.  {\bf 9} (1983)  302.
\bibitem{myrev}  S.~V.~Ketov,  Int.\ J.\ Mod.\ Phys.\ A {\bf 28} (2013) 1330021, arXiv:1201.2239 [hep-th]. 
\bibitem{hi1} F. Bezrukov and M. Shaposhnikov, Phys. Lett. B 659, 703 (2008), arXiv:0710.3755 [hep-th].
\bibitem{hi2} F. Bezrukov, D. Gorbunov and M. Shaposhnikov, JCAP 0906 (2009) 029, arXiv:0812.3622 [hep-ph].
\bibitem{ks} S.V.~Ketov and A.A.~Starobinsky, JCAP {\bf 08} (2012) 022, arXiv:1203.0805 [hep-th].
\bibitem{planck2} P.~A.~R.~Ade {\it et al.}  [Planck Collaboration], Astron. Astrophys. {\bf 571} (2014) A22;  arXiv:1303.5082 [astro-ph.CO].
\bibitem{bicep2}  P.~A.~R.~Ade {\it et al.}  [BICEP2 Collaboration], Phys. Rev. Lett. {\bf 112} (2014)241101;  arXiv:1403.3985 [astro-ph.CO].
\bibitem{kw11} S.~V.~Ketov and N.~Watanabe,  Phys. Lett. {\bf B741} (2015) 242; arXiv:1410.3557 [hep-th].
\bibitem{des} B.~J.~Broy, F.~G.~Pedro and A.~Westphal, JHEP {\bf 1503} (2015) 03, 029; arXiv:1411.6010 [hep-th].
\bibitem{planck3} P.~A.~R.~Ade {\it et al.}  [Planck Collaboration], {\it Planck 2015 results: XIII. Cosmological parameters}; arXiv:1502.01589 [astro-ph.CO].
\bibitem{kt1} S.~V.~Ketov and T.~Terada, Phys. Lett. {\bf B736} (2014) 272: arXiv:1406.0252 [hep-th].
\bibitem{kt2} S.~V.~Ketov and T.~Terada, JHEP {\bf 1412} (2014) 062; arXiv:1408.6524 [hep-th].
\bibitem{ktalk} S.~V.~Ketov, {\it Natural inflation and universal hypermultiplet}, arXiv:1402.0627 [hep-th].
\bibitem{kte} S.~V.~Ketov and T.~Terada, JHEP 1312 (2013) 040, arXiv:1309.7494 [hep-th].
\bibitem{lwtr} B.~Whitt, Phys. Lett. {\bf B145} (1984) 17.
\bibitem{kkw}  S.~Kaneda, S.~V.~Ketov and N.~Watanabe, Mod. Phys. Lett.
{\bf A25} (2010) 2753, arXiv:1001.5118 [hep-th].
\bibitem{nat1} K.~Freese, J.~A.~Frieman and A.~V.~Olinto, Phys. Rev. Lett. {\bf 65} (1990) 3233. 
\bibitem{nat2} F.~C. Adams, J.~R.~Bond, K.~Freese, J.~A.~Frieman and A.~V.~Olinto, Phys. Rev. {\bf D 47} (1993)  426, 
[hep-ph/9207245].
\bibitem{nat3} K.~Freese and W.~H.~Kinney, JCAP {\bf 1503} (2015) 044; arXiv:1403.5277 [astro-ph.CO].
\bibitem{smo} A.~Spencer-Smith, {\it  Higgs vacuum stability in a mass-dependent renormalization scheme}, 
arXiv:1405.1975 [hep-ph].
\bibitem{vag}  R.~Myrzakulov, L.~Sebastiani and S.~Vagnozzi, Eur. Phys. J. {\bf C75} (2015) 444;
arXiv:1504.07984 [gr-qc].
\bibitem{book} S.~V.~Ketov, {\it Quantum Non-linear Sigma-models}, Springer-Verlag, Heidelberg, 2000.
\bibitem{bd} N.~D.~Birrell and P.~C.~W. Davies, {\it Quantum Fields in Curved Space}, Cambridge Univ. Press, New York,
1982.
\bibitem{bos} I.~L.~Buchbinder, S.~D.~Odintsov and I.~L.~Shapiro, {\it Effective action in Quantum Gravity}, Taylor
and Francis, New York, 1992. 
\bibitem{lal} Z. Lalak, D. Langlois, S. Pokorski and K. Turzynski, JCAP {\bf 0707} (2007) 014, arXiv:0704.0212 [hep-th].
\bibitem{jap} A.~A.~Starobinsky, S.~Tsujikawa and J.~Jokoyama, Nucl. Phys. {\bf 610} (2001) 383, arXiv:astro-ph/0107555. 
\bibitem{ktsu} S.~V.~Ketov and S.~Tsujikawa, Phys. Rev. {\bf D86} (2012) 023529, arXiv:1205.2918 [hep-th].
\bibitem{uk} C. van de Bruck and M.~Robinson, JCAP {\bf 1408} (2014) 024; arXiv:1404.7806 [astro-ph.CO].
\bibitem{uk2} C. van de Bruck and L.~E. Paduraru, {\it The simplest extension of Starobinsky inflation}, arXiv:1505.01727 [astro-ph.CO].
\bibitem{k2} R.~Greenwood, D.~Kaiser, and E.~Sfakianakis, Phys Rev {\bf D87} (2013) 064021, 
arXiv:1210.8190 [hep-ph].
\bibitem{mukh} V.~Mukhanov, {\it Physical Foundations of Cosmology}, Cambridge Univ. Press, 2005, 421 pages.
\bibitem{sp} D.~Polarski and A.~A.~Starobinsky, Nucl. Phys. {\bf B385} (1992) 623.
\bibitem{klin1} R. Kallosh and A. Linde, {\it Non-minimal inflationary attractors}, arXiv:1307.7938 [hep-th].
\bibitem{klin2} R. Kallosh and A. Linde, {\it Multi-field conformal cosmological attractors}, arXiv:1309.2015 [hep-th]. 
\bibitem{d1} D.~I.~Kaiser, E.~A.~Mazenc and E.~I.~Sfakianakis, Phys. Rev. {\bf D87} (2013) 064004; 
arXiv:1210.7487 [astro-ph.CO].
\bibitem{dsprl} D. Kaiser and E. Sfakianakis, Phys. Rev. Lett. {\bf 112} (2014) 011302; arXiv:1304.0363 [astro-ph.CO].
\bibitem{planck13} P.~A.~R.~Ade {\it et al.}  [Planck Collaboration], Astron. Astrophys. {\bf 571} (2014) A15;  arXiv:1303.5075 [astro-ph.CO].

\end{thebibliography}
\end{document}